# Theoretical evaluation of electrochemical cell architectures using cation intercalation electrodes for desalination

Kyle C. Smith*

*Department of Mechanical Science and Engineering,*
*Computational Science and Engineering Program,*
*Beckman Institute,*
*University of Illinois at Urbana-Champaign*
*Urbana, IL 61801, USA*

*Corresponding author's email: kcsmith@illinois.edu

**Abstract**

Water scarcity is a dilemma facing much of the global population. Cation intercalation desalination (CID) cells, which use intercalation host compounds (IHCs) in combination with ion-exchange membranes (IEMs), could aid in addressing this challenge by treating saline water sources. Originally, the performance of such cells was predicted utilizing continuous flow of saline water through porous IHC electrodes. Here, we use two-dimensional porous-electrode theory with concentrated solution transport to evaluate the performance of various cell architectures where flow occurs through open flow channels (OFCs) when two IHC electrodes comprised of nickel hexacyanoferrate (NiHCF) are used to store $Na^+$ ions. We show that, when two OFCs are used, cation exchange membranes (CEMs) are adjoined at flow-channel/electrode interfaces, and an anion exchange membrane (AEM) is arranged between flow channels, salt removal increases relative to the original design with flow-through (FT) electrodes. The IEM stacking sequence within such a membrane flow-by (MFB) cell is the fundamental repeat unit for electrodialysis (ED) stacks using many IEMs (CEM/AEM/…/CEM/AEM/CEM) with many diluate streams. Accordingly, we simulate the performance of such ED stacks using NiHCF IHCs, and we predict that salt adsorption capacity (per unit NiHCF mass) is amplified by twenty-fold relative to MFB and FT cells, while simultaneously decreasing 0.7 M NaCl feed water to 0.2-0.3 M within diluate streams. The generality of these findings is further supported by simulations using $Na_{0.44}MnO_2$ IHC instead of NiHCF. Thus, we propose the use of cation IHCs as alternatives to the gas-evolution reactions used in conventional ED.

1. Introduction

Recent analysis shows that the majority of the world's population experiences water scarcity for at least one month of the year.[1] Water reuse and desalination of salt-rich water sources (e.g., sea and brackish water) could reduce the burden of freshwater scarcity.[2] Pressure-driven reverse osmosis technology has substantial installed capacity around the world[2,3] but requires large-scale plants to desalinate water efficiently.[4] Alternative membrane technologies exist to desalinate water using electric potential as a driving force. Electrodialysis (ED) is the most developed of such technologies and has found extensive use in demineralization of salt-containing solutions.[5] In ED Faradaic reactions are used to induce electric potential drop across a stack of ion-exchange membranes (IEMs) with alternating selectivity toward cations and anions. When saline source water is pumped through flow channels between IEMs their selectivity enables the generation of alternating streams of concentrated brine and desalted water, referred to as concentrate and diluate respectively.[6] Conventional ED stacks use gas-evolution reactions (e.g., $H_2$ and $O_2$ gases[7]) to generate ionic current, and as a result costly metals[8] and large stacks are required.

Solid and solution-phase electrode processes offer benefits over the gas-evolution reactions used in conventional ED. Along these lines, reactions involving iron-based redox couples in solution, including hexacyanoferrates anions ($Fe(CN)_6^{4-}/Fe(CN)_6^{3-}$), have been evaluated for use in reverse ED,[9–11] but their performance may be limited by crossover through IEMs due to their mobility in solution (as is commonly encountered in flow batteries using dissolved redox couples[12]). In contrast, capacitive deionization (CDI) uses the electric double-layers (EDLs) of high surface-area porous carbon to store cations and anions in solution.[13] CDI cells have also been developed with IEMs arranged on the surface of electrodes (MCDI), so as to minimize co-ion expulsion within EDLs.[14] Capacitive electrodes have also been incorporated into reverse ED to increase energy recovery from salinity gradients.[15] Other efforts in the CDI literature have been aimed at increasing salt removal, including through the use of flow electrodes[16] and a hybrid arrangement of Na-ion and capacitive electrodes.[17,18] Novel bi-porous carbons have also been employed to enable flow of electrolyte through the thickness of electrodes, rather than along the electrode's length.[19]

To enable desalination of seawater-level salt concentrations, other devices have employed solid-state Faradaic electrode reactions in lieu of capacitive electrodes. Specifically, the desalination battery used a Na-ion intercalation cathode paired with a Ag/AgCl conversion anode.[20] More recently, we predicted that a Na-ion battery containing intercalation electrodes can desalinate seawater-level salt concentrations if an anion exchange membrane (AEM) is used to suppress Na-ion transport between the electrodes.[21,22]

This concept, which we refer to presently as cation intercalation desalination (CID), can be employed with generic intercalation host compounds. In our original work,[21,22] we applied the NID concept with $Na_{0.44}MnO_2$ (NMO) and $NaTi_2(PO_4)_3$ (NTP), which exhibit sizable volumetric charge capacities (approximately 200 mAh/mL-NMO[23] and 400 mAh/mL-NTP[24]). Despite these advantages, the abuse tolerance of these materials may be limited due to the degradation of NMO as a result of over (dis)charge[25] and the propensity of NTP to hydrolyze in moderate pH solutions.[24,26] In contrast, the open framework structure of Prussian Blue Analogues (PBAs) has enabled facile intercalation and long cycle life in various aqueous cation batteries (including $Na^+$,[27–30] $K^+$,[27,28] $Ca^{2+}$,[31] and $Zn^{2+}$,[32] among other ions in general electrochemical cells[33]). While Prussian Blue itself ($Fe_2(CN)_6$) is soluble in aqueous solution, its analogous structures formed by substitution of one Fe atom within $Fe_2(CN)_6$ with a different transition metal, including Ni,[27] Cu,[28] and Mn[34] are insoluble in aqueous and non-aqueous electrolytes. Thus, PBAs are metal hexacyanoferrate compounds, which use the same redox-active unit as hexacyanoferrate ionic complexes,[9–11] but can be used as solid-state IHCs in aqueous electrolytes. Because of their open-framework structure, PBAs show substantially lower volumetric charge capacity than NMO and NTP (approximately 100 mAh/mL-PBA[27]). Aside from energy storage, nickel hexacyanoferrate (NiHCF) and copper hexacyanoferrate PBAs have recently been used to harvest energy from salinity gradients in aqueous NaCl solutions.[35,36] In addition, these materials can be synthesized in Na-rich[29,30] and Na-deficient[27,28] forms, enabling the construction of symmetric CID cells.[21]

Theoretical and computational modeling is being employed increasingly to guide the development of electrochemical desalination devices, including in ED,[37–46] capacitive deionization,[14,19,47–51] CID,[21,22] and other technologies.[52–55] Computational implementations of these models vary in dimensional fidelity, including zero-,[39–42,49,51] one-,[19,46] and two-dimensional models.[21,22,37,38,43–45,48,55] While capacitive[19,50,51] and Na-ion[21,22] models capture the local dynamics of charge adsorption, previous ED models have focused on accurate description of transport processes within flow channels and membranes,[37–44] while neglecting the Faradaic reactions that draw current and induce electric field.

In this work we use a numerical model to predict the performance of Na-ion desalination cells with various membrane and flow arrangements. We model nickel hexacyanoferrate as an intercalation host compound and show that, despite its low charge capacity, efficient desalination of seawater-level concentrations is possible in a range of CID device configurations. To perform these simulations we extend the fidelity of our two-dimensional electrochemical model by including concentrated solution effects as well as IEMs with ideal permselectivity. We show that electrodialysis stacks using Na-ion intercalation electrodes can desalinate large volumes of water efficiently when optimized flow configurations are employed.

## 2. Modeled System

In our evaluation of cell architectures we consider designs with various numbers of IEMs, extending from the flow-through type CID cells simulated previously, up to ED stacks with many

IEMs. Figure 1a depicts an ED stack using NiHCF intercalation electrodes simulated here. The stack is designed with a streamwise length $L$ of 20 mm, porous electrodes with thickness $w_e$ of 1.0 mm, and infinitestimally thick ion-exchange membranes (IEMs) spaced apart by a channel thickness $w_c$ of 0.4 mm. Within the cell an alternating series of cation-exchange membranes (CEMs) and anion-exchange membranes (AEMs) are used to induce production of diluate and concentrate from influent saltwater flowed with a steady, fully developed cross-plane velocity $v(x)$. A cell voltage $V_{cell}$ is measured between the stack's positive and negative terminals as a constant total current $I$ is applied to the positive terminal.

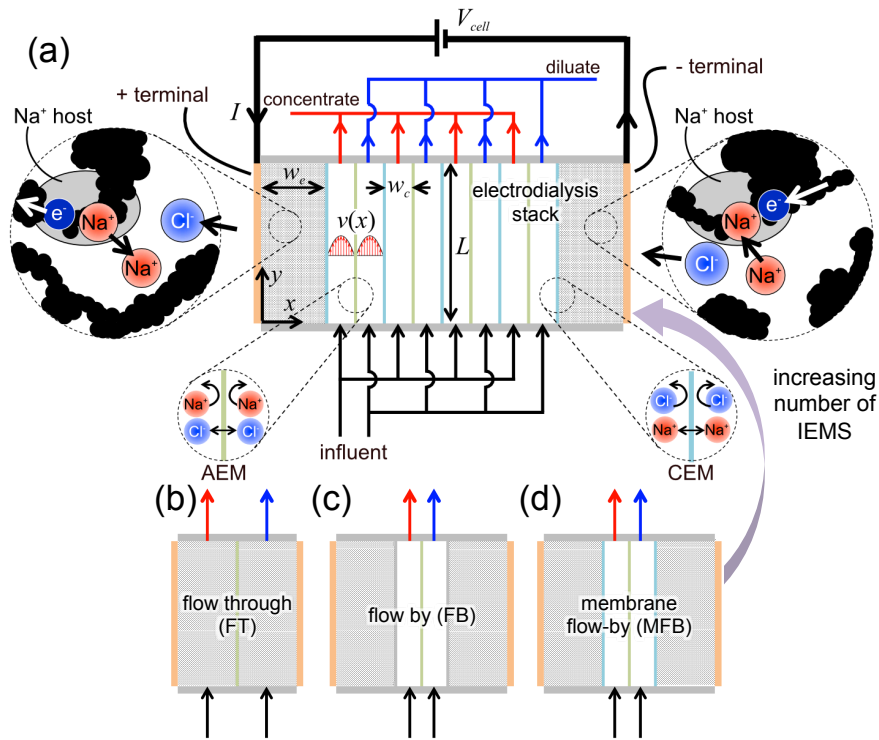

**Figure 1**: (a) Schematic of a simulated electrodialysis stack utilizing Na-ion intercalation electrodes. Several other cells with only two flow channels were also simulated with Na-ion intercalation electrodes, including (b) a flow-through (FT) cell, (c) a flow-by (FB) cell, and (d) a membrane flow-by (MFB) cell.

Several distinct designs using only two flow channels are included in the set of cell architectures that we evaluate. In the first cell (Fig. 1b) "flow-through" porous intercalation electrodes separated by an AEM (as in Ref. 21). We then introduce an open flow channel in between each electrode and the AEM, so as to "flow by" the electrodes (Fig. 1c). Finally, we introduce a CEM at each electrode/flow-channel interface, producing a "membrane flow-by" cell (Fig. 1d), which is the shortest possible electrodialysis stack (i.e., because it has only two flow channels and three membranes).

A porous-electrode model is used here to simulate ionic conduction, salt diffusion, membrane polarization and transport, intercalation reactions, and electronic conduction for each cell architecture. The governing equations and boundary conditions for the latter two processes have been described previously,[21] and we omit their description here. Presently we enhance the modeling of ionic conduction and salt diffusion by incorporating concentrated-solution effects for aqueous NaCl, and we implement IEMs with generic cation transference numbers.

We modify the current-conservation equation for aqueous NaCl solution to include the so-called thermodynamic factor $\gamma_\pm$ accounting for concentrated solution activity:[56]

$$\nabla \cdot \left( -\kappa_{eff} \left( \nabla \phi_e - \frac{2R_g T}{F}(1-t_+)\gamma_\pm \nabla \ln c_e \right) \right) - a v_s i_n = 0 \ , \tag{1}$$

where $\phi_e$, $c_e$, and $t_+$ are respectively solution-phase potential, salt concentration, and the transference number of cations in solution, and $R_g T/F$ takes its usual meaning. Ionic current density within the electrolyte appears as the argument of the divergence operator in Eq. 1, $\vec{i}_e = -\kappa_{eff}\left(\nabla \phi_e - 2R_g T/F(1-t_+)\gamma_\pm \nabla \ln c_e\right)$. The source term couples the intercalation current density $i_n$ (given by reaction kinetics described in Ref. 21) of intercalation host particles loaded at volume fraction $v_s$ to ionic current density in solution $\vec{i}_e$, where $a$ is the volumetric surface area of intercalation host particles. Experimental data[57] for the mean-molar activity coefficient $f_\pm$ as a function of $c_e$ was used to compute $\gamma_\pm$ as $\gamma_\pm = (1 + \partial \ln f_\pm / \partial \ln c_e)$.[56] The effective ionic conductivity $\kappa_{eff}$ is approximated by Bruggeman theory in terms of the bulk solution-phase ionic conductivity $\kappa$ as $\kappa_{eff} = \varepsilon^{1.5} \kappa$, where $\varepsilon$ is porosity. To model IEMs with arbitrary cation transference number $t_{m,+}$ we include diffusion potential and Donnan potential to determine the solution phase potential drop from side $i$ to side $j$ of a given membrane:

$$\phi_{e,i} - \phi_{e,j} = \frac{2R_g T}{F}(1 - t_{m,+}) \ln\left(\frac{f_{\pm,i} c_{e,i}}{f_{\pm,j} c_{e,j}}\right). \tag{2}$$

Equation 2 can be obtained from an expression for electrostatic membrane potential[58] using the definition of solution-phase potential as the reduced electrochemical potential of Na ions in solution (see Refs. 21,22,56). We enforce the membrane transference number $t_{m,+}$ by constraining the fraction of cationic current in solution to $\vec{i}_{e,+} = t_{m,+} \vec{i}_e$, where the cationic current density is $\vec{i}_{e,+} = -t_+ \kappa_{eff} \nabla \phi_e$ and $\vec{i}_e$ is the solution-phase current density appearing as the argument of the divergence operator in Eq. 1. Here, we assume ideally permselective AEMs and CEMs with $t_{m,+} \approx 0$ and $t_{m,+} \approx 1$, respectively. In our treatment of IEMs we also neglect water transport through them, as well as their resistance. In reality it is known that IEM permselectivity and resistance are coupled and depend on thickness, swelling,[59] and solution concentration.[60,61] Though incorporating detailed IEM properties is clearly important to obtain agreement between theoretical predications and experimental results when using particular IEMs, we consider ideally permselective IEMs to focus on the "best case" scenario of IEM performance among all cell architectures that we consider. For more details we refer the reader to other recent studies that do incorporate these effects.[38]

In addition we modify the salt conservation equation to include the experimentally measured[62] bulk chemical diffusion coefficient of salt $\tilde{D}$:

$$\frac{\partial(\varepsilon c_e)}{\partial t} + \nabla \cdot (\vec{v}_s c_e) + \nabla \cdot (-\tilde{D}_{eff} \nabla c_e) - a v_s (1-t_+) \frac{i_n}{F} = 0 \tag{3}$$

where the effective chemical diffusion coefficient of salt is given by Bruggeman theory as $\tilde{D}_{eff} = \varepsilon^{1.5} \tilde{D}$. We calculate bulk ion conductivity $\kappa$ using the following equality for concentrated binary electrolytes with complete salt dissociation:[56] $\kappa = c_e F^2 \tilde{D} / (2 R_g T \gamma_\pm t_+ (1-t_+))$. Using this approach the present model explicitly captures the polarization due to ohmic conduction and concentration boundary layers within salt water in electrodes and flow channels. We note that the present approach to modeling ionic transport in the electrolyte by salt conservation (Eq. 3) and ionic current conservation (see Eq. 2) equations builds on the so-called "porous electrode theory" developed by Newman,[56,63,64] which has been applied readily to Li-ion batteries. Though, in the dilute limit, this approach is equivalent to the typical Nernst-Planck formulation used in CDI modeling,[48,50,54] which expresses individual-species flux and invokes their conservation individually, it is limited to the modeling of binary electrolytes.

We model the (de)intercalation of Na-ions within NiHCF according to the following reaction:[30]

$$x_{Na} \text{Na}^+ + x_{Na} e^- + \text{NaNiFe(CN)}_6 \rightleftarrows \text{Na}_{1+x_{Na}} \text{NiFe(CN)}_6, \tag{4}$$

where the stoichiometric factor $x_{Na}$ is the fraction of intercalated Na. We initialize the positive electrode in the Na-rich state with $x_{Na}$ = 99.958% and the negative electrode in a sodium-deficient state with $x_{Na}$ = 0.042%. Since NiHCF has been synthesized in either Na-rich[30] or Na-deficient[27] forms, reduction or oxidation of either compound would be required to prepare the two electrodes with this initial stoichiometry. The equilibrium potential of intercalation $\phi_{eq}$ is approximated by that of a regular solution of adatoms and vacancies with negligible pair interaction energy:[65]

$$\phi_{eq} = \phi_{eq}^0 + \frac{R_g T}{F} \ln\left(\frac{1-x_{Na}}{x_{Na}}\right), \tag{5}$$

where the reference potential $\phi_{eq}^0$ is approximated as 0.60 V vs. SHE by the 50% state-of-charge potential measured from galvanostatic charge/discharge at low C-rate in 1 M $Na_2SO_4$.[27] As shown in Fig. 2, this potential model agrees very well with the experimental data of a large range of $x_{Na}$ with the largest deviation of approximately 40 mV near $x_{Na}$=5%.

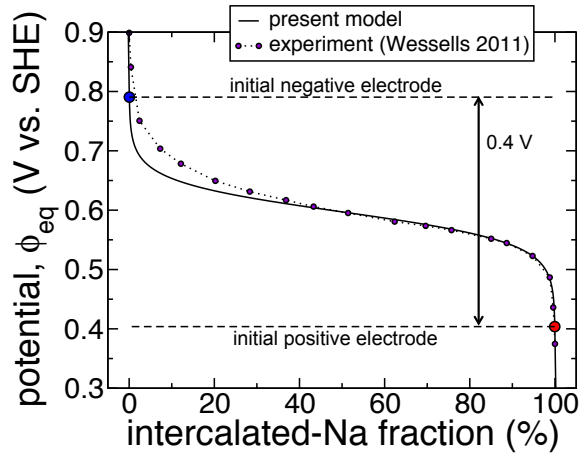

**Figure 2**: Modeled equilibrium potential as a function of intercalated-Na fraction for the presently simulated Prussian Blue Analogue. The initial states chosen for both positive and negative electrodes are shown, producing an initial open-circuit cell voltage of 0.4 V. Experimental data for low-rate cycling in 1 M $Na_2SO_4$ are shown for comparison.[27]

We neglect the kinetic polarization and mass transfer resistance within NiHCF particles because of their small size and high rate capability.[27,30] In practice, we employ Butler-Volmer kinetics (as described in Ref. [21]) with a finite rate constant of $2\times10^{-11}$ mol/m²-s per (mol/m³)$^{1.5}$ (similar to that of NMO[21]) and with the volumetric surface area of 50 nm particles. For these conditions we find that order of magnitude changes in rate constant affect polarization by less than 5 mV at 1C.

The terminal concentration of intercalated Na inferred from ideal stoichiometry (Eq. 4) and X-ray diffraction over-estimates the charge capacity of NiHCF by approximately 50% due to the presence of interstitial water.[27,30] Consequently, we scale the terminal concentration of intercalated Na from the ideal, theoretical value of 6.26 M to 4.10 M, so as to match observed capacities at low cycling rate (approximately 60 mAh/g[27,30]). The porous electrodes simulated here include 50 vol.% NiHCF and 40 vol.% porosity.

Several cells using Na-ion intercalation compounds are simulated to evaluate scale-up of NID to electrodialysis stacks. In each case common dimensions are used for cell designs (1.0 mm electrode thickness, 0.4 mm membrane spacing, and 20 mm channel length). Each cell is charged and discharged at 54 A/m$^2$ average current density with a volumetric flow rate through each flow channel of 32 mL/hr per unit meter of cell depth. For the electrodes simulated here containing 50 vol.% NiHCF and 40 vol.% porosity, this current density corresponds to a theoretical charge/discharge time of 1 hr (i.e., a C-rate of 1C). For the present current density and flow conditions 505mM salt removal is expected by application of Faraday's Law,[†] which is 72.1% salt removal for the 700mM influent simulated here. In practice charge/discharge processes are terminated if cell voltage reached a certain value, as in our previous work.[21] We terminate charge/discharge if cell-voltage reaches either -0.488 V, -0.577 V, -0.754 V, -0.931 V, or -1.107 V for stacks having 2, 4, 8, 12, or 16 flow channels, respectively, to accommodate the high polarization in large ED stacks. In addition, charge/discharge processes are also terminated if the local reduction potential within either electrode reaches 3.21 V vs. Na$^+$/Na$^0$, which is slightly below the O$_2$ evolution potential at 1 M NaCl concentration and neutral pH, so as to prevent O$_2$ evolution.

## 3. Results and discussion

In this work we perform a systematic evaluation of CID cell architectures ranging between a cell using two intercalation electrodes to desalinate water from one diluate stream (originally reported in Ref. 21) and an electrodialysis stack using two intercalation electrodes to desalinate water from many diluate streams simultaneously. We begin by evaluating cells that desalinate only one stream at time with a given pair of electrodes.

We consider three such cells: (1) a flow-through (FT) design in which influent is pumped through the microscopic pores of electrodes separated by an AEM (Fig. 1b), (2) a flow-by (FB) design in which influent is pumped through an open channel between the electrode and an AEM (Fig. 1c), and (3) a membrane flow-by (MFB) design similar to the former, except that, additionally, a CEM is arranged at the interfaces between the electrodes and the influent streams (Fig. 1d).

Aside from our interest in scaling CID cells in a systematic manner, there are a number of technological reasons for comparing performance among these cells. Specifically, the FT cell poses mechanical design challenges, because of the low fluidic permeability of cast electrodes with small pores. Alternatively, a (FB) design can be employed where influent water flows in open channels instead of through porous electrodes. Furthermore, membranes are commonly employed in FB CDI, called membrane CDI or MCDI, to reduce energy consumption.[66] Accordingly, we evaluate the performance of FB cells using CEMs adjacent to electrodes in an MFB cell configuration.

We first examine the effect of these flow/IEM configurations on cell voltage, effluent salinity, and the spatial distribution of salt within these cells. Figure 3a shows the variation of cell voltage as a function of charge time (i.e., negative time corresponds to discharge) for the first charge/discharge cycle of these three NID cells. The MFB cell shows improved capacity and similar polarization to the FT cell, despite the larger thru-plane thickness of the MFB cell. In contrast, when the FB cell is used without CEMs lesser capacity is obtained and more polarization is incurred than with CEMs (i.e., when an MFB cell is used). The capacity obtained during charge and discharge is affected by polarization because each half cycle is terminated at a certain cutoff voltage (see Methods section). As a result, though all cells here utilize the same electrodes with the same theoretical capacity, the discharge utilization for each cell will be different due to the magnitude of the polarization that they incur during cycling. Figure 3b shows effluent salinity for these cells as well, in all cases averaged across the outlet of the corresponding flow-channel exit. All three cells eventually produce effluent near theoretical levels, but the FB cell requires nearly three-fold longer time to reach the desired effluent salinity than the FT and MFB cells.

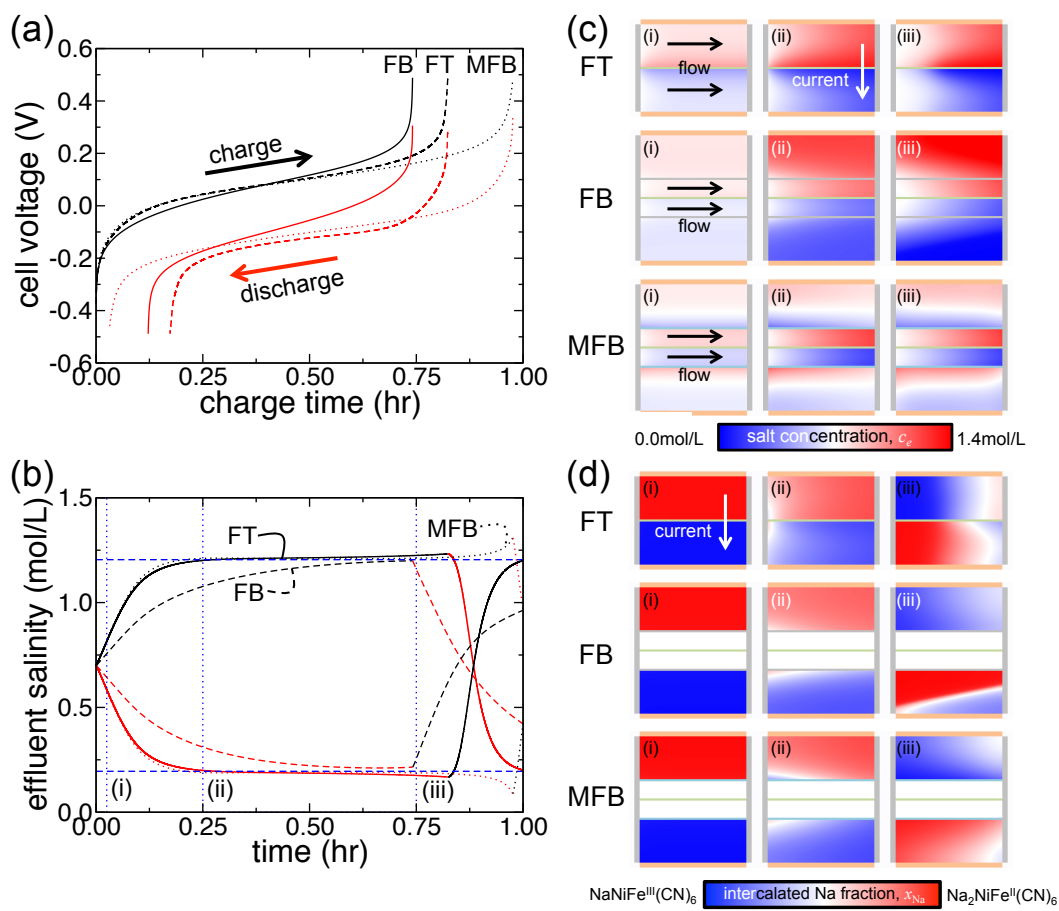

**Figure 3**: Simulated cycling characteristics for flow-through (FT), flow-by (FB), and membrane flow-by (MFB) cells. (a) Cell voltage as a function of time is shown for both charge (black curves) and discharge steps (red curves) of the first cycle. (b) Effluent salinity as a function of time is shown for both concentrate (black curves) and diluate (red curves) during the first cycle. Theoretical effluent concentrations are shown as blue dashed lines. (c) Snapshots of salt concentration and (d) intercalated-Na fraction for three instants in time are marked in (b).

The aforementioned results reveal that the MFB cell configuration is preferable to the FB configuration without CEMs. The superior performance of the MFB cell is a result of its ability to retain salt within Na-ion electrodes. This phenomenon is revealed in Fig. 3c, which shows snapshots of salt concentration within each cell at several instants during charging. Though after 30 minutes of charging the FB cell achieves theoretical desalination levels, salt concentration within the positive electrode decreases to the same degree due to salt transport at the electrode/flow-channel interface. In contrast, salt concentration within the electrodes of the MFB cell is non-uniform due to concentration polarization at the CEM, but on average it remains at the initial salinity level. Figure 3d shows the spatial variation in composition of NiHCF particles within the electrodes, represented here as intercalated-Na fraction, at the same instants during charging. FT and MFB cells show streamwise propagation of an intercalation reaction zone (as observed previously[21]) because of the IEM polarization at the electrode edge. In contrast, in the FB cell intercalation reactions progress through the thickness of electrodes because conventional separators (that are not selective) small polarization.

The MFB cell exhibits several benefits over the other two architectures including its flowability, high capacity, and low polarization. The structure of alternating IEMs within the MFB cell (CEM/AEM/ … /CEM) is similar to that found in ED stacks, the main difference being that the MFB cell contains two flow passages while ED stacks can theoretically employ any multiple of two flow channels.

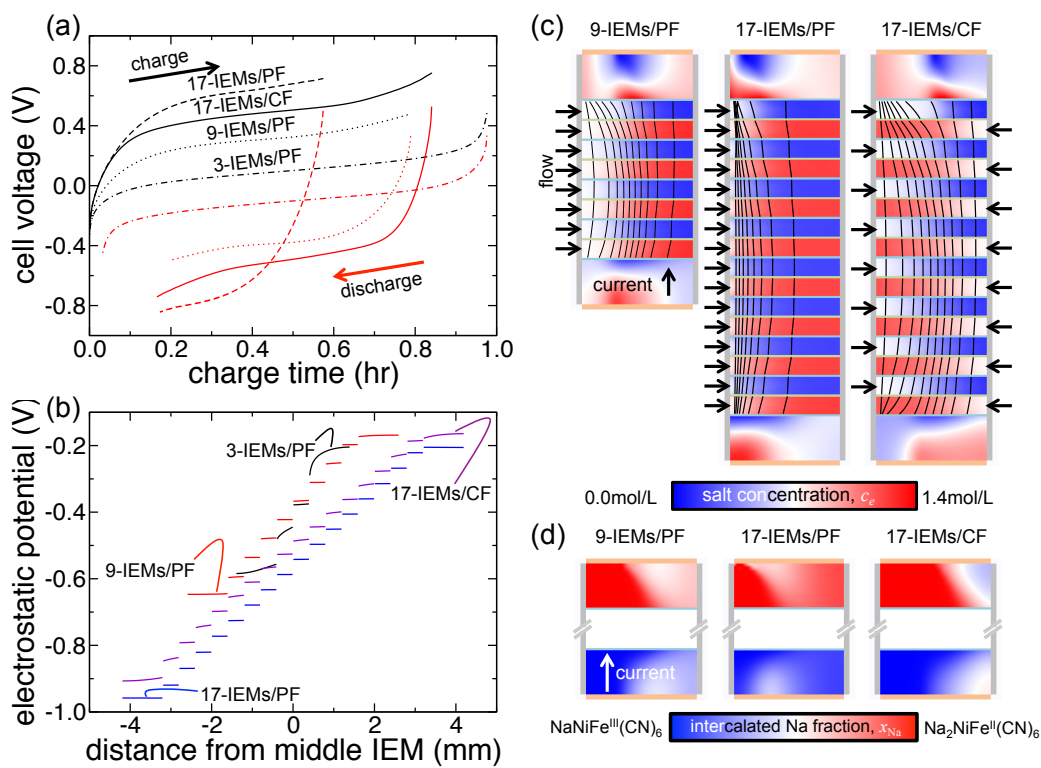

**Figure 4**: Simulated cycling characteristics for electrodialysis stacks using two NiHCF electrodes with 9 IEMs and parallel flow (PF), 17 IEMs and PF, and 17 IEMs and counterflow (CF). (a) Cell voltage as a function of time is shown for both charge (black curves) and discharge steps (red curves) of the first cycle. (b) Variation of electrostatic potential with distance from the middle IEM near the outlet of each cell at the end of the discharge cycle. (c) Snapshots of salt concentration and (d) intercalated-Na fraction at the end of discharge are shown for each cell. Solution-phase current-density lines (black lines) are overlaid on the salt concentration fields within the flow channels.

Accordingly, we now explore the effect of adding more flow channels to the MFB cell, introducing a novel ED stack that uses Na-ion intercalation electrodes. Shown in Fig. 4 are the cycling characteristics of ED stacks with 9 and 17 IEMs comprising 8 and 16 flow channels, respectively. In addition, two different flow scenarios were tested: (1) parallel flow (PF) where influent in adjacent channels flows in the same direction and (2) counterflow (CF) where influent flows in opposing directions. We note that alternative flow modes and stack configurations have been considered in electrodialysis previously.[67] Upon examination of the voltage curves for these stacks (Fig. 4a), we observe that average polarization increases as the number of IEMs increases (or, equivalently, flow channels). To understand this effect the electrostatic potential profile between the two electrodes was examined at $x = L$ (Fig. 4b). This profile shows that the electrostatic potential drop also increases substantially with the number of IEMs, an effect which is primarily due to IEM polarization.

But the electrostatic potential distribution, alone, does not explain the variations in capacity utilization observed among each of the cases. Under the present conditions the 17-IEM cell in CF achieves 67% utilization, in spite of its increased polarization relative to the 3-IEM cell. In contrast, the 17-IEM cell in parallel flow attains only 40% utilization. To understand the origin of the observed capacity variations, we examine the solution-phase current and salt concentration distribution within each stack. Figure 4c shows salt concentration at the end of discharge and the lines along which ionic current flows within the electrolyte in the flow channels. Figure 4d shows the distribution of intercalated Na within NiHCF particles in the electrodes at the same instants in time. Current-density lines (similar to heat-flux lines or "adiabats" in conduction heat transfer) can be understood alternatively as surfaces through which no current flows. Further, the position of these current-density lines has been chosen here such that 10% of the total current flows within the solution between adjacent lines. We note that few previous models[45] of ED have examined current distribution within stacks. For the 9-IEM, PF cell these lines are uniformly spaced, revealing that ionic current flows uniformly and flows along the shortest path from the positive to the negative electrode. When the number of flow channels is doubled to 16 by using 17 IEMs, the current distribution in PF becomes focused at the inlet of the flow channels where current-density lines are grouped closely. The high current density at the inlet is responsible for the higher polarization and lower capacity produced by this stack. This behavior is due to the non-uniform IEM polarization that occurs in PF. In contrast, when a CF arrangement is used uniform current distribution is produced even with 17 IEMs. Here, uniform current density is enabled by CF's greater uniformity of IEM polarization. We note that despite the variations in capacity and the non-uniformity of current density we observe that each of these cells produce effluent near the theoretical salinity levels.

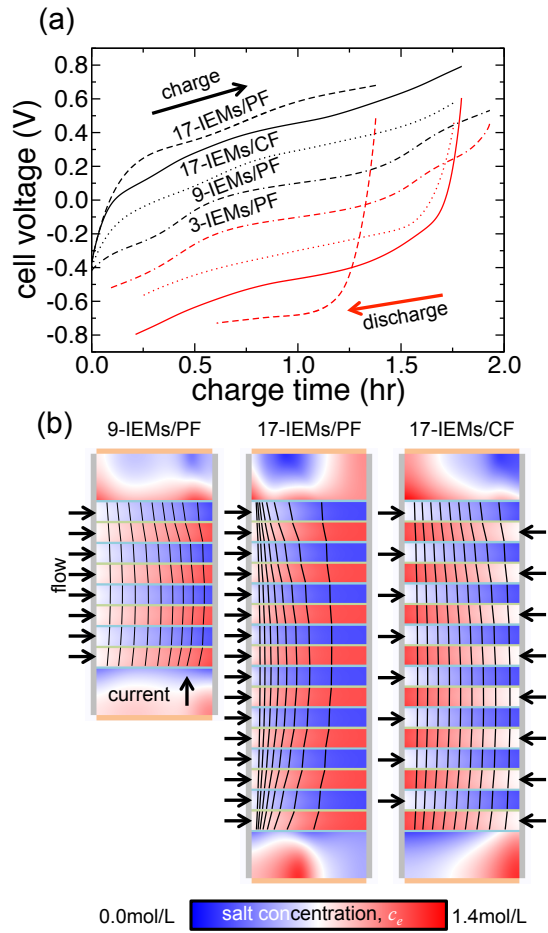

**Figure 5**: Simulated cycling characteristics for electrodialysis stacks using two NMO electrodes with 9 IEMs and parallel flow (PF), 17 IEMs and PF, and 17 IEMs and counterflow (CF). (a) Cell voltage as a function of time is shown for both charge (black curves) and discharge steps (red curves) of the first cycle. (b) Snapshots of salt concentration at the end of discharge are shown

for each cell. Solution-phase current-density lines (black lines) are overlaid on the salt concentration fields within the flow channels.

To determine whether these effects are particular to NiHCF electrodes, we also simulated results for ED stacks using NMO electrodes with the NMO-specific model parameters described in Ref. 21. As shown in Fig. 5a, the trends of capacity and polarization are similar to those observed with NiHCF electrodes, but NMO cells cycle for substantially longer times (when cycled at the same current density and with the same volumetric loading of the intercalation host compound) as NiHCF as a result of NMO's larger charge capacity in comparison with NiHCF. We also observe that the NMO-based cells are capable of achieving the same degree of salt removal as NiHCF-based cells (Fig. 5b).

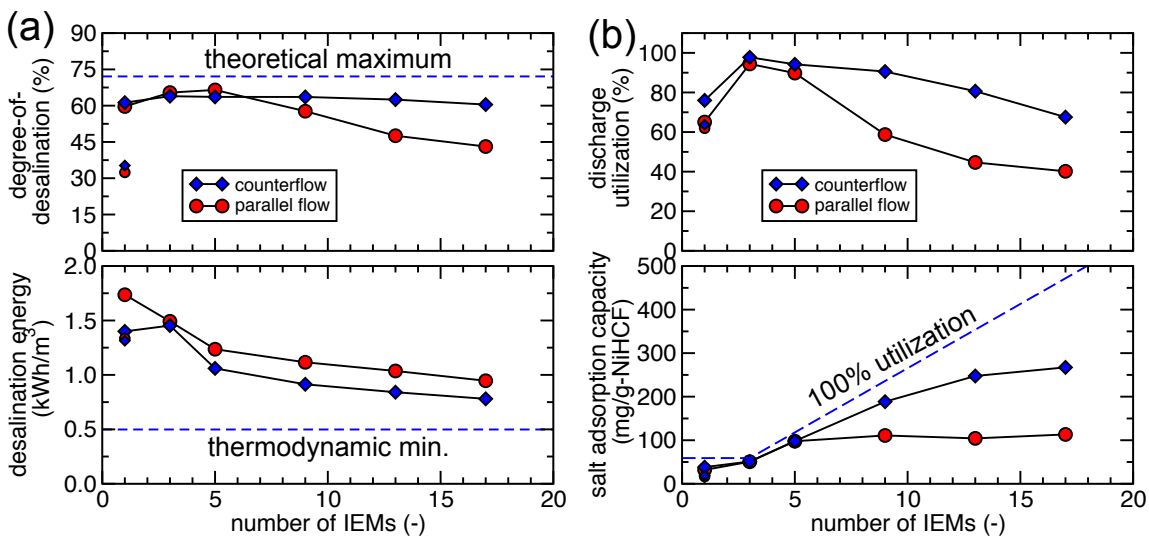

**Figure 6**: (a) Degree-of-desalination and desalination energy and (b) discharge utilization and salt adsorption capacity as a function of number of IEMs. All data are shown for the discharge step of the first cycle. For cells with one IEM, data for the FB cell (without CEMs) are shown

separately with small symbols, while that of the FT cell is plotted as part of the discrete series of ED stacks with large symbols.

We calculated desalination metrics for ED stacks constructed with various numbers of IEMs using only two NiHCF intercalation electrodes. Figure 6a shows the degree-of-desalination (salt removal normalized to influent salinity of 0.7 M) with a range of cells: FT with one AEM, FB with one AEM and no CEMs, MFB with one AEM and two CEMs, and a range of ED stacks. The results for both CF and PF configurations are shown for each cell. All CF cases except FB show salt removal in excess of 60%. In contrast, PF exhibits reduced degree-of-desalination for large stacks due to their non-uniform current distributions (see Fig. 4c). Figure 6a also shows the energy consumed per cubic meter of diluate produced (referred to as desalination energy) assuming no energy is recovered during charge/discharge processes. In general, desalination energy decreases with increasing stack size, and it approaches the thermodynamic minimum energy consumption[‡] within 60%. This decreasing trend occurs because ohmic and kinetic losses within the electrodes are distributed over a larger volume of product effluent for large stacks.

An advantage of incorporating Na-ion electrodes into an ED stack is the ability to remove large amounts of salt with small amounts of intercalation host compound. If 100% utilization of intercalation capacity is assumed, the salt adsorption capacity (SAC) of the NiHCF electrodes increases with the size of ED stacks (Fig. 6b). SAC, a metric commonly used in the CDI literature, is defined here as the mass of NaCl desalinated per unit mass of NiHCF. As shown in Fig. 6b, simulated utilization is less than 100% due to the combined effects of IEM polarization and flow-channel resistance. When the MFB cell is operated in CF its SAC of 51 mg/g approaches the theoretical limit (59 mg/g), while FT and FB cells produce only 38 mg/g and 18 mg/g, respectively. Large stacks show the most pronounced effect of flow configuration on diluate production. 17-IEM cells produce 267 mg/g in CF and 113 mg/g in PF, respectively. Even greater utilization, salt adsorption capacity, and stack scaling may be achieved by recirculation of electrolyte through flow channels and electrodes.

**4. Conclusions**

We have predicted that Na-ion intercalation compounds can be used as efficient and reversible electrodes for electrodialysis. We consider a nickel-hexacyanoferrate Prussian Blue Analogue, which, despite its low charge storage capacity (100 mAh/mL-PBA[27]), is capable of continuously removing 0.4-0.5 M NaCl from 0.7 M influent. We observe that the distribution of ionic current within flow channels is biased toward the inlet when concentrate and diluate streams flow in the same direction (i.e., parallel-flow configuration), but these effects can be mitigated by flowing concentrate and diluate streams in opposing directions (i.e., in counterflow configuration). Additional strategies could be used to further enhance performance, including the use of forced convection of electrolyte within intercalation electrodes and recirculation of electrolyte through flow channels. The energy consumption per unit diluate volume decreases as the number of membranes or flow channels is increased in the stack, showing the promise of using Na-ion electrodes for efficient electrodialysis. The present porous-electrode model explicitly captures two-dimensional ion transport within all flow channels of the electrodialysis stacks considered. In contrast, many previous two-dimensional electrodialysis models begin with a periodically repeating domain[37,38,43,44] that cannot capture the asymmetric current distributions that we observe here. Accordingly, the present results suggest that similar non-uniform current distributions could manifest in conventional electrodialysis, and stack performance could be enhanced by using the counterflow arrangement that we propose. Furthermore, Na-ion electrodialysis stacks could be used in *reverse electrodialysis* to harvest energy from salinity gradients introduced by flowing influent of dissimilar salt concentrations into the stack. We note, also, that the impact of non-ideal IEM transport is not yet quantified and, therefore, deserves attention. We also note that the demonstration of nickel hexacyanoferrate electrodes in desalination applications has not yet been reported in the literature and such experiments will enable further research development in the emerging research area of cation intercalation desalination.

**Footnotes**

†An upper bound on salt removal can be estimated by applying Faraday's law to the desalinating electrode. The resulting salt removal is $c_e^{in} - \bar{c}_e^{out} = iL/wFu_s$,[21,22] where $c_e^{in}$ and $\bar{c}_e^{out}$ are influent salinity and average diluate salinity, respectively, and $i$ and $u_s$ are the average applied current density and the electrolyte velocity, respectively.

‡We compute the thermodynamic minimum energy as the reversible work of separation $W_{rev,sep}$ using the concentrated solution activity model described in the text:

$$W_{rev,sep} = R_g T \left( 2(V_d + V_c) c_e^{in} \ln(f_\pm^{in} c_e^{in}) - V_d c_{e,d}^{out} \ln(f_{\pm,d}^{out} c_{e,d}^{out}) - V_c c_{e,c}^{out} \ln(f_{\pm,c}^{out} c_{e,c}^{out}) \right).$$

Here, the subscripts c and d denote, respectively, diluate and concentrate and $V$ is effluent volume. For the present results with 50% water recovery and theoretical effluent concentrations from Faraday's Law, the thermodynamic minimum energy is 0.50 kWh/m³-diluate.